## Developing Mathematical Creativity with Physics Invention Tasks

*Suzanne White Brahmia, University of Washington*
*Andrew Boudreaux, Western Washington State University*
*Stephen E. Kanim, New Mexico State University*

**Abstract**

Modeling the creative mathematical sensemaking that characterizes expert thinking in physics is typically a struggle for new learners. To help students learn to reason this way, we created a set of supplemental activities called *Physics Invention Tasks* (PITs). PITs engage students in *quantification,* the process of mathematically generating quantities central to modeling in physics. PITs depict a scenario in which students need to engage in decision-making associated with creating quantities in order to resolve a problem. Students make decisions about the arithmetic construction, the magnitude and associated unit, and in some cases spatial direction of a new quantity. The design of PITs is informed by *Inventing with Contrasting Cases*, an instructional approach shown to foster *generativity,* i.e., creativity with mathematical structures. In this paper, we describe the theoretical foundation of PITs, and their structure and implementation. We share preliminary observations of the impact of PITs in one course, averaged over two years, as measured by the FCI and CLASS. We see an improvement in the FCI normalized gain corresponding to the introduction of PITS and we see CLASS pre- to post-test gains that may be the highest reported from a large enrollment calculus-based physics course. We discuss future research into the learning mechanisms and instructor influence associated with PITs.

## I        Introduction

Physics instructors' customized algebraic reasoning is a struggle for students to emulate in introductory physics courses.[1-3] Experts' conceptualization of arithmetic operations and of symbolic representations facilitates the visualization of the relationships between physically meaningful quantities. In introductory physics, models are generally quantitative, and rely on reasoning from students' prior mathematics courses. These prerequisite courses primarily focus on developing procedural expertise and axiomatic reasoning independent of context (*i.e.,* using pure numbers). As such, their learning objectives do not necessarily align with the ways that physicists reason mathematically with physical quantities. Students' struggles with the mathematics in their physics courses may have less to do with procedural difficulties than with *mathematizing* the physical world, *i.e.* engaging in mathematical sense-making with the quantities that underpin the modeling of phenomena. Developing mathematical reasoning in physics contexts, which include $\sim 10^2$ unfamiliar quantities, is a physics education problem. We suggest that all students can benefit from instruction that helps them refine cognitive structures in the context of new quantities, such as proportional reasoning and invariant reasoning, which are hallmarks of mathematizing physics.

The instructional design of Inventing with Contrasting Cases has been shown to help pre-college students better understand the deep structures of invariant, composite quantities in several contexts (e.g., density, statistical variance).[4-7] We thus view ICC as a promising approach for helping to deepen college students' understanding of the product, ratio, difference and net quantities that underpin mathematical modeling in introductory physics. This paper presents Physics Invention Tasks, a novel set of ICC-based, supplemental instructional activities. PITs aim to develop mathematical reasoning by providing opportunities for students to quantify the physical world.

As a preliminary investigation of the impact of PITs, we present data collected in a large-enrollment, calculus-based mechanics course using two common measures: the *Force Concept Inventory* (FCI)[8] and the *Colorado Learning Attitudes about Science Survey* (CLASS).[9] The introduction of PITs into the course curriculum was accompanied by improvement in FCI gain scores and also by positive novice-to-expert shifts in categories of the CLASS associated with mathematical attitudes and beliefs. These results suggest that use of PITs in interactive engagement physics courses may help achieve learning outcomes associated with mathematization.

In section II, we describe the theoretical foundations for PITs. Section III describes the structure and implementation of PITs, while section IV presents evidence of the impact of PITs on student learning and attitudes. The conclusion, section V, discusses questions for further research.

## II        Theoretical Foundations for Physics Invention Tasks

In this section, we describe the framing that underpins both the *instructional work that PITs do* and *how they do it*. PITs are designed to help students learn to mathematize in physics, and Conceptual Blending Theory (CBT) is a framework that helps us understand expert physics mathematization as a subconscious inclination.  The instructional design if Inventing with Contrasting Cases provides a pedagogical structure intended to help students develop subconscious mathematical inclinations.

### *Mathematization and CBT*

Expert-like mathematization in physics involves both procedural and conceptual mastery of the mathematics involved. [10, 11] Gray and Tall highlight the distinction between these facets of understanding, explaining that "the symbol 3/4 stands for both the process of division and the concept of fraction."[10] They define *proceptual* understanding, in which *proc*edural mastery and con*cept*ual understanding coexist, as an appropriate goal for instruction.  A student with a proceptual understanding of fractions, for example, would move fluidly between the procedure of dividing 3 by 4, and the physical instantiation of the fraction ¾ as a precise quantification of portion. Similarly, a physics student with a proceptual understanding of momentum would move fluidly between the procedure, multiplying a scalar mass by a vector velocity, and conceptualizing the product $m\vec{v}$ as a quantity unto itself (*i.e.*, as simply **p**), with its own, emergent properties and utility.

We suggest that viewing physics and math as separate spheres of thinking may impede proceptual development in physics, by leading some students to separate "doing math" from "doing physics." Rebello *et al*. found that students typically did not spontaneously draw on their trigonometry and calculus knowledge to solve problems in a physics context. [12]  Rowland reports that engineering majors enrolled in a differential equations course struggled to make sense of the units of kinematics quantities they were familiar with, observing that *"few students were able to determine the units of a proportionality factor in a simple equation."* [13]  When designing instruction, even at the level of introductory physics, we find it productive to consider the physical and mathematical worlds as inseparable.

Conceptual blending theory (CBT) provides a framework for understanding the integration of mathematical and physical reasoning. [14]  In their theory, Fauconnier and Turner describe a cognitive process in which a unique mental space is formed from two (or more) separate mental spaces. The blended space can be thought of as a product of the input spaces, rather than a separable sum.  According to CBT, development of expert mathematization in physics would occur not through a simple addition of new elements (physics quantities) to an existing cognitive structure (arithmetic), but rather through the creation of a new and independent cognitive space.  This space, in which creative, quantitative analysis of physical phenomena can occur, involves a continuous interdependence of thinking about the mathematical and physical worlds.

For most students, establishing a blended thinking space of this type requires cultivation. Physics Invention Tasks can help students tightly integrate mathematical and physical reasoning, opening pathways for thinking generatively with mathematics in physics contexts. Developing models in physics depends in part on proceptual facility with two important mental habits: 1) *quantification,* the process of generating new quantities to characterize properties of a system, and

2) *seeking invariance,* finding quantities that maintain the same value in a variety of situations.[4, 15] Figure 1 illustrates an arithmetic reasoning blend, in which two distinct domains of thinking are merged to form a new cognitive space optimally suited for quantification and seeking invariance.

Prior research has identified quantification as a significant challenge to students who are learning to mathematize.[16, 17] Math educator Patrick Thompson defines quantification as "the process of conceptualizing an object and an attribute of it so that the attribute has a unit of measure, and the attribute's measure entails a proportional relationship…with its unit." For example, a train's motion can be quantified by a *momentum* relative to the earth, which combines the mathematical objects of ratio, product and vector. Thompson considers quantification to be "a root of mathematical thinking," and argues that learners develop their mathematics from reasoning about quantities. In work with middle school algebra students, Ellis claims that modes of mathematical structural reasoning are more likely to develop when students practice with "emergent" quantities, rather than the strictly numerical patterns and algorithms common to school mathematics.[18] Ellis uses the term *emergent* to mean that the quantity is composed of other quantities through multiplication or division (e.g., speed); we avoid this term because of its temporal connotation in physics, and refer instead to *composite physical quantities*. Ellis claims it is precisely these kinds of quantities that help develop students' abilities to create powerful generalizations.

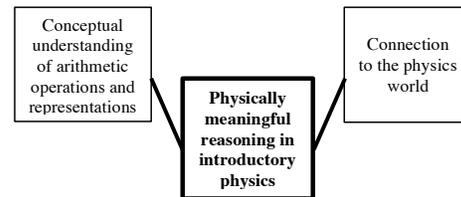

Fig. 1. Arithmetic reasoning about the physical world as a conceptual blend.

To illustrate how reasoning with quantities enriches cognition, consider students' initial framing of multiplication as a process of repeated addition. This framing is productive in the context of pure numbers, and is relatively simple for new learners to visualize. Contrast this with momentum: as a product of mass and velocity, momentum can be understood only by framing multiplication as an operation that produces something entirely different from its factors – a composite physical quantity. A yearlong introductory course presents $\sim 10^2$ new physical quantities; precisely because of this focus on modelling with quantity, physics instruction has an important role to play in helping students develop proceptual understandings not only of physics but also of mathematics.

Physics quantification is the first step in modeling how systems do and don't change. Learning physics quantification involves developing the mental habit of searching for useful invariant quantities, and developing a sense about which quantities are likely to be invariant. It is in this context that we refer to "seeking invariance", which maps to the mathematical notion of *invariant under transformation*. At the introductory level, invariant quantities may be intrinsic properties of matter (e.g., density, specific heat) or systems (e.g., energy, momentum.) Mathematics education researchers consider understanding of multiplicative invariance to be an essential part of understanding ratios,[19] which are ubiquitous in physics both for quantification and for making comparisons. PITs are designed on the assumption that students can benefit from cultivating the habit of seeking invariance. In the remainder of this section we describe how quantification and seeking invariance are woven into the learning design of Inventing with Contrasting Cases.

### Inventing with Contrasting Cases

The use of contrasting cases is at the heart of quantification and mathematical modeling in physics. Imagine, for example, a physicist who would like to model the cars' motion during a car race. She begins by quantifying the rate at which the cars speed up and makes the assumption that the rate of increase in speed is independent of speed – that it is an invariant of the motion. To check the validity of the assumption, she could measure the speed changes during various intervals of a single car's motion, and compare the ratio of the change in speed to the elapsed time for these intervals. She might then use this new quantity to compare between cars, i.e. contrast cases of data collected from different cars' motions.

PITs are informed by the theoretical underpinnings of Inventing with Contrasting Cases (ICC), a learning design refined by mathematics education researcher Dan Schwartz and colleagues.[4-7] ICC scaffolds the process described above by constraining the experimental design, and the assumptions for the students. The students then invent an invariant as a means of characterizing a system. [4] Materials are designed to help students learn to notice information they might otherwise overlook, and learn to ignore information that is not helpful for the characterization being made.

Figure 2 illustrates the contrasting cases used in a task in which students invent a "speeding up index" task developed by Schwartz, Chase, Oppezzo and Chin).[4] Stacked sketches depict contrasting data sets, with simple artifacts (drops of oil that fall from the cars at a uniform rate) that students can use for measurement. Students decide that cars with a same index value should speed up by the same amount in identical time intervals. They also recognize that the initial speed and car type are irrelevant.

Fig. 2. Sketch included with the Speeding up index invention task.

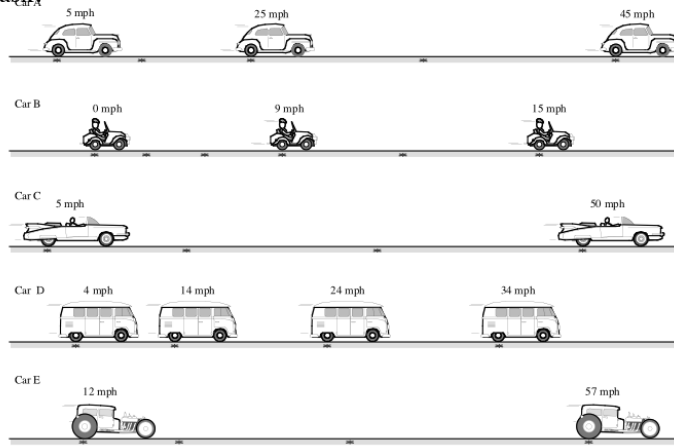

PITs involve *productive failure*, in which learners are encouraged to persist in generating methods for approaching novel tasks. [20] ICC generally involves substantial trial and error thinking, and a student group may fail to generate the canonical solution. This is expected; the instructional value is hidden in the sense that students are primed to subsequently quickly recognize how and why the canonical solution is the most sensible response to the challenge at hand. [7, 20] To maximize the productivity of failure, scaffolding can be adjusted to seek balance between unproductive struggle and excessive prescription. [21, 22] During the speeding up index task, students struggle with the unfamiliar units (mph per oil drop), and with interpolation. While those struggles may prevent initial success, they bring students closer to notions of constant rate of change and to meaningful interpretation of the units of acceleration ($m/s^2$).

To create a productive and intellectually safe place for students to be creative, PITs are founded on a social-constructivist view of learning. [23, 24] The mathematical constructions both challenge and guide individual reasoning, yet the successes and failures are mediated by the group and not by the individual. The sharing of failure bolsters students' willingness to take risks and builds a shared understanding that being wrong in physics is part of eventually being right.

Research-validated social constructivist learning designs are not new to physics education (for a summary, see Meltzer and Thornton[25]). PITs differ from other activities in that they target developing physics creativity at a level of generating new physical quantities and with the intention of developing an appreciation for the deep structure of an invariant and its quantification. They are also intended to prime students for formal instruction, not replace it, and have no expectation that all students will arrive at a canonical solution. Students struggle to generate mathematical statements from scratch using only arithmetic operations, and this struggle itself is the learning objective of an invention task. Unlike modeling curricula,[26] where students develop whole algebraic models, PITs focus at a finer grain size by asking students to invent the elements of those models – composite physical quantities that characterize nature.

## III     Implementing Physics Invention Tasks

We begin this section with a detailed description of one task, involving a *Car Washing Inefficiency Index* (Fig. 3), which we use at the beginning of a sequence of tasks leading to students' first product quantity - the formal concept of work. We then describe structural features common across most PITs.

### Detailed Example

We have developed a sequence of three invention tasks to precede formal instruction on the concept of work. Work is known to be challenging for students:  Lindsey, Heron & Shaffer found that students commonly confound work with force, [27] and Brahmia & Boudreaux report the more general result that students commonly attribute to a scalar product the vector properties of its factors.[1]  The entire Work sequence is found on the PIT website.[28]  Here we describe in detail the first task in the sequence, in which students invent a Car Washing Inefficiency Index. The task is based on the context shown in Fig. 3.

This is the first invention task students encounter that targets creating a product index. Students consider three car wash outlets, each with two teams of workers. For each team, students are shown the number of workers and the time in minutes required to wash a Toyota Camry.  The challenge is to create an *inefficiency index*, with a larger value indicating a more inefficient team. (The index is motivated by a storyline in which management is deciding which outlet is in most urgent need of further training.)  A constraint is introduced in order to highlight contrasting cases: students are told that teams at the same car wash outlet must have the same index value.

We find that students (as well as physics faculty in professional development workshops) initially reach for ratio as an invariant measure when they are confronted with this new challenge, a sensible step since ratios have been useful thus far. The target index in this case, however, is the *product* of the number of workers on the team and the number of minutes required to wash the car.

Many students are quick to propose the ratio of the number of minutes to the number of persons on the team.  Typical conversations involve an interpretation of this ratio as "the number of minutes per person."  Groups will often agree that this is a good measure of relative inefficiency. They then find, however, that index values for teams at the same car wash do not match, in violation of the externally imposed constraint.   This realization generally induces cognitive dissonance and a return to sensemaking.  We consider this to be an example of *productive failure* as it creates a need to consider a new approach:  product-as-measure.

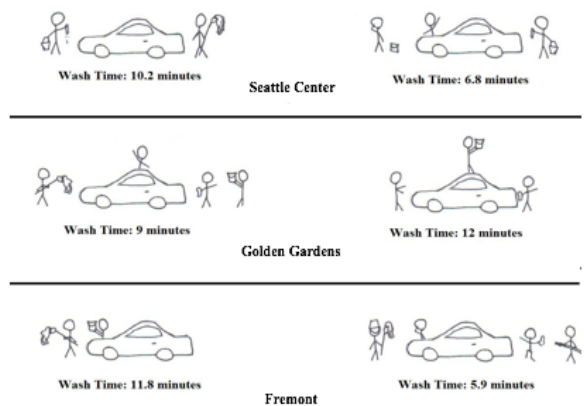

Fig. 3. Sketch presented to students as part of the *Car washing inefficiency index task*.

### Structure

Figure 4 illustrates the interplay of the theoretical foundations and the learning trajectory students typically follow during a physics invention task. PITs are founded on Harel's Necessity Principle: [29]   *"For students to learn what we intend to teach them, they must have a need for it, where 'need' refers to intellectual need, not social or economic need."* The "mission" of a PIT (e.g.

to characterize inefficiency) establishes an intellectual need, and requires that students be creative with arithmetic operations. The novelty of choosing their own mathematical operations makes many students uncomfortable at first – perhaps because most physics students are accustomed to having a specific target procedure to learn. The need to complete the mission validates the more generative interaction with mathematics required by an invention task.

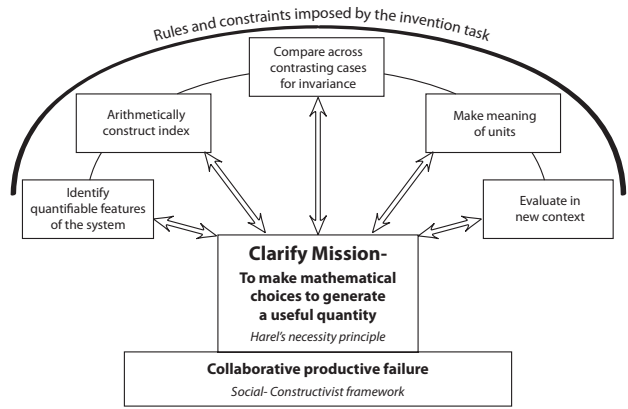

Fig. 4: Student progression in a typical PIT

To make progress toward understanding the mission, students first identify quantifiable features of the system and combine them in a variety of ways. The drawings provided with the task are primitive, involving only a limited number of quantifiable features. The level of scaffolding is designed to provide sufficient framing,[21,22] while also giving students practice with the commonly missing first step of modeling, as described by Redish and Smith[30] : "We [experts] begin … by choosing a physical system we want to describe. … we have to decide what characteristics of the system to pay attention to and what to ignore. This is a crucial step and is where much of the skill or "art" in using math in STEM lies." As the students become more comfortable with invention, the tasks may include more non-essential features, requiring students both to decide how to use the information provided, and to select relevant features. Next, the groups typically try out some arithmetic operations, and ponder whether their ideas make sense, frequently returning to the mission. This interplay deepens their interpretation of the new quantity. Testing out various operations continues until the group reaches consensus on their index. They test their index with the contrasting cases to ensure that a single procedure works for all of these cases. This step is essential in helping to develop the inclination for seeking invariance.[4]

Once students are satisfied with their index, they proceed to the follow-up questions. For each task, groups are asked about the units of their index. In the case of a ratio quantity they are asked to describe the meaning of the value of the unit rate in the context of its units. For example, in the speeding up index (Fig.2) the students are asked "Use everyday language to describe the specific information that the speeding up index tells you about the car's motion," where the anticipated response is that it tells us how much the speed changes for each oil drip. As a product quantity, the interpretation of the units of the inefficiency index is more open-ended. This index is measured in person-minutes, which can be interpreted as the number of minutes it would take one worker to wash a car or the number of workers it would take to wash a car in one minute. Unlike the interpretation of the units of a ratio quantity, there are many valid interpretations of these units. The analogous discussions of subsequently encountered product quantities (e.g. work, momentum, torque) draws on this example. The emphasis on units here is intended to establish the cognitive link between value and unit that is essential to the understanding of *quantity*.[15] In addition, the groups are asked questions involving scaling and proportional reasoning, usually with whole number factors to promote practice of mathematical reasoning using simple-valued physical quantities. The remaining follow-up questions vary depending on the known physics-specific difficulties from the physics education research literature.

We have used invention tasks in a variety of settings, including large-enrollment as well as studio courses, and courses for science and engineering majors as well as general education courses for non-science majors. In some cases we combine tasks into a sequence (e.g. work), while in other cases a single task can stand-alone, depending on the complexity of the quantity or its mathematical structure (see Appendix for typical in-course contexts.). We find that physics invention tasks are, in general, accessible and engaging to students in all of these situations. As part of the social-constructivist design, students collaborate in small peer groups. To allow for productive failure, we limit intervention to situations in which the group reaches an unproductive standstill and is getting frustrated. We avoid guiding students to the normative reasoning at this stage. We generally provide closure, after small group work, by leading a full group discussion. The discussion provides a segue and link from the students' own sense-making and generativity around the challenge presented by the context of the invention task to recognizing the utility of the physics quantity in subsequent formal instruction.

Invention tasks associated with a variety of physical quantities can be found online (~30 tasks at the time of this writing).[28] Included are instructor notes describing implementation and sequencing in a typical introductory physics progression. We emphasize that invention tasks are designed to complement existing course curriculum. We allow 20-40 minutes of class time per sequence initially, but as students become more efficient with inventing, we find that about 20 minutes suffices, generally done as we introduce a challenging new physical quantity, perhaps with one sequence every two weeks. Inevitably, developing deep, proceptual understanding requires sacrificing some breadth of coverage. We advocate strongly for this trade-off. Because invention work primes students to understand subsequent formal instruction, some "lost" time is subsequently recouped and, we argue, a deeper understanding of physical quantities is gained.

## IV. Impact of Physics Inventions Tasks on student learning

PITs were piloted at Rutgers University in *Extended Analytical Physics* (EAP),[31] a large-enrollment, introductory calculus-based course designed for mathematically underprepared engineering students. The preliminary measures of the impacts on student learning and attitudes presented below were made in this context. In order to characterize learning, we present pre/post comparisons using two measures in common use in physics education - the Force Concept Inventory (FCI), an established measure of conceptual understanding of Newtonian mechanics, and the CLASS-physics, currently the most commonly used measure of student attitudes and beliefs about physics learning. We make an historical comparison using the FCI, and an international comparison with reported results using the CLASS.

*Context*

EAP is a calculus-based course into which students are enrolled if they are concurrently enrolled in pre-calculus rather than calculus in their first semester.[31, 32] EAP is offered as an alternate to Analytical Physics I (API), the mainstream calculus-based course, and the student population of EAP differs demographically from API. Between 40%-60% in a given year of the EAP students are in the Educational Opportunity Fund program, which provides financial and other support services to first-generation, economically disadvantaged students. The portions of the student population from under-represented minority groups and the portion that is female are both typically larger in EAP than in API, 12% and 21% in API respectively in 2013. (see Table I for comparison.)

**Table I:** Demographics of EAP Fall 2013

| # of students | 110 |
|---|---|
| *Mean Mathematics SAT (2013 test)* | 610 |
| *% African American or Latino* | 40% |
| *% female* | 29% |

EAP uses an adaptation of The Investigative Science Learning Environment (ISLE), which provides a setting for students to engage in science practices to construct physics ideas.[33] We note that because EAP lacks a lab, we do not consider it a full-blown ISLE course. PITs form a regular part of the curriculum, which can be thought of as a hybrid of abridged ISLE infused with PITs. We emphasize that PITs are an add-before activity and do not represent a curriculum on their own.

*Research methods*

The FCI was administered as a pre-and post-test in years both before and after PITs were implemented as a regular curricular activity. We select 2003 as a comparison year because it is the most recent year for which we have pre-PITs FCI data. Between 2003 and 2013 there were no significant changes made to the course aside from the introduction of PITs; there was a small decrease in the students' mean SAT math score, which we have ignored in this analysis, as its only small effect would be to strengthen our claims. The lecturer was the same in all three years, but the teaching assistants were not. Between 2003 and 2013 the Institutional Research Board requirements changed, which resulted in a smaller percentage of student data being available for the post-PIT condition (currently consent forms are required, and in 2003 they weren't.) We therefore combined the first two years during which PITS were fully implemented (2013 and 2014) to form the comparison cohort of the course using PITs.

In the years under comparison, the FCI pretest was administered as an ungraded quiz under exam conditions at the beginning of the semester. The students were not constrained by time and they were awarded credit for completion. The students took the post test, administered under the same conditions, during the second-to-last week of the semester. In the years after PITs were implemented the students took the CLASS online (outside of class) as a pretest during the first week of class, and as a posttest during the last week of class. Completion of the in-class FCI in addition to completing the CLASS online surveys increased a student's overall course grade by 2%.

We report on the CLASS after the introduction of PITS only, as we don't have CLASS data available before the implementation of PITS. The dataset represents matched pre/post results of the students who opted in to have their data used in our study, and we have eliminated any CLASS surveys on which students did not select answer 4 for question 31 (an indicator of not taking the test seriously). Due to opting out, not taking one of the surveys or lazy survey completion, our matched data set represents ~60% of the students.

Comparing end-of-semester course grades (on a 100-point scale) of the students represented in the FCI and CLASS post-PITS samples to those not in the samples, and adjusting for the students' grades that did not earn the 2% bonus, we see no statistically significant difference in the average course grade between the students in the groups of students reported on here and those not reported on. We consider the samples to be representative of the class as a whole.

*Results*

The FCI provides a cognitive measure, and students' pre-to-post "gain" provides one type of learning measure which allows for a comparison of course gains before and after the introduction of PITs. Figure 5 compares FCI scores of a cohort for whom PITs had been integrated as a regular part of instruction to a baseline measurement taken in EAP prior to the introduction of the PITs (from 2003, n=102). The number of students for whom we have matched pre/post and who agreed to allow their results to be part of this study is n=144 (from 2013/14 combined). The mean scores in 2003, before PITs, are pre = 40%±1.6% and post = 61%±1.5%, where the uncertainty is the standard error. The mean scores of 2013 and 2014 combined, after implementation of PITs, are pre = 36%±1.3%, and post = 66%±1.4%.

The course normalized gain is the ratio of what was gained to what could have been gained, (post – pre)/ (100%-pre) and is a commonly reported measure. [34] The normalized gain of the course averages using matched samples in 2003 is 0.35 ± 06 and in 2013 is 0.47 ± 05, which are both in the range of interactive engagement courses. The trend shows an improvement in learning as measured by the FCI. Comparing internationally across institutions, a robust result on the CLASS is a significant shift *away from* expert-like attitudes and beliefs about physics over the course of calculus-based introductory physics, even in courses using research-validated approaches that measure large conceptual gains.[9] A "good" CLASS result is to "do no harm" and have a neutral shift; there are few examples of excellent results from large-enrollment calculus-based courses in which the shifts are positive.[35] In this study, we observe a neutral overall shift. We observed neutral or positive shifts in most specific categories (the overall shift includes questions that do not cluster into categories.).

**Figure 5:** FCI comparison (before the introduction of PITs, 2003, n=102) and after (2013/14, n=144)

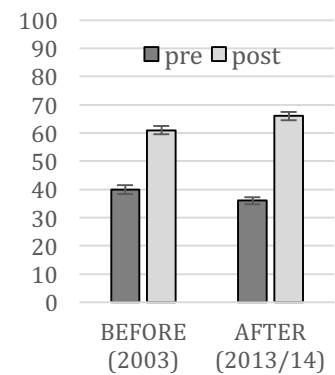

To address more specifically the impact of PITs on students' attitudes about mathematical reasoning, we focus here on the categories associated with students' attitudes and beliefs about the use of mathematics in physics. Figure 6 shows the categories that specifically address mathematical reasoning in physics (e.g. an expert response is to disagree with the statement "I do not expect physics equations to help my understanding of the ideas, they are just for doing calculations.") The pre-to-post comparison in each of these five categories has a small positive effect size (Cohen's d = +0.2) and is significant at the 95% confidence interval using a two-tailed repeated t-test of significance. The results for all categories, effect sizes and p-values are shown in the Appendix. While slightly negative-to-neutral CLASS gains in these categories are characteristic of ISLE in the large enrollment algebra-based course at Rutgers, the problem solving categorical gains seen here have not been reported on in any large enrollment ISLE-based course. We attribute the positive CLASS categorical gains to the modified ISLE curriculum enriched by the PITs.

*Discussion*

In this preliminary investigation, we've shown that PITs can result in increased cognitive gains as measured by the FCI, and can impact students' attitudes and beliefs about the role of mathematics in physics reasoning in a large-enrollment course. The gains seen on the problem-solving categories of the CLASS are comparable to those reported in highly successful, but lower enrollment, calculus-based courses with a diverse student population using MI (Modeling Instruction).[36] This similarity with MI makes sense; we see PITS as similar in that students are collaboratively generating symbolic descriptions, but different in that PITs target generating mathematical structure at the level of physical quantities. In addition, the published MI positive CLASS results are all from courses that have small enrollments (<30 students);[36] the result reported here are from a large-enrollment lecture course. We believe the results reported here are the highest reported CLASS problem solving gains in a calculus-based course with an enrollment >100 students.[35]

We interpret these results as potential indicators of a shift in the learning culture catalyzed by PITs, both for students and for instructors, which leads to more expert-like attitudes and beliefs about equations in physics. We observe that through collaborative productive failure, students' cognitive struggle with algebraic reasoning using abstract quantities becomes validated as being both challenging and fruitful for learning. We observe instructors become more aware of the cognitively blended world that situates their own physics reasoning, and that their students' struggle is not about lack of ability to do algebra. We believe that instructors commonly re-examine their own thinking about the interplay between arithmetic sense-making and physical quantities, and devise their own ways of helping their students develop a conceptual mathematical understanding. We suggest that PITs can facilitate better impedance matching between learners and instructors in the context of mathematics in physics.

**Figure 6:** CLASS- physics categories associated with mathematical reasoning, pre-instruction and the gains over one semester. Sample is combined from Fall 2013 and Fall 2014, n=121. The error bars represent the standard error.

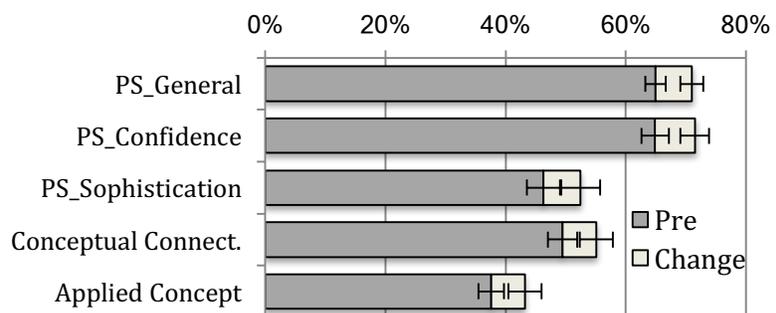

While we have shown that PITs can have a positive effect on student learning, the measures used are not explanatory of the mechanism leading to improved conceptual gains and more expert-like mathematical attitudes - they are not direct measures of productive mathematical sensemaking. The strongest statement we can make is that including PITs in an interactive engagement course improved students' performance on the FCI, and had a positive impact on students' attitudes and beliefs about mathematical reasoning as measured by the CLASS.

## V.    Conclusion

We present PITs as a set of carefully designed collaborative activities that engage students in the novel practice of mathematical creativity. We argue that the process of inventing quantities to characterize the physical world on an as-needed basis is precisely the kind of mathematical reasoning that characterizes physics thinking and that, at a fundamental level, PITs engage students in the creative mathematical reasoning of experts.

PITs are designed to help both students and instructors to better focus on the key features of seeking invariance and quantification as a routine part of mathematical reasoning in introductory physics. PITs have been field-tested in the context of supplementing already transformed courses that routinely involve research-based collaborative learning methods and materials. We emphasize that PITs are designed as an enhancement, and are not intended to be a stand-alone transformation of a traditionally taught college classroom.

As a comparison with other similar transformed calculus-based courses that do not use ICC, this paper describes a study using measures that are commonly reported on in the literature: 1) an historical comparison using a conceptual cognitive measure (FCI) and, 2) an international comparison using an affective measure of students' attitudes and beliefs about the role of mathematics in physics (CLASS – problem solving categories).  The PITs vs no PITs FCI comparison has a small positive effect size. We believe that the gains for the CLASS problem solving categories are the largest published gains for courses with enrollment >100.[35] We interpret these result as showing promise for PITs to help improve mathematical reasoning as part of physics learning and to help render student attitudes and beliefs more expert-like about the role of mathematics in calculus-based physics in courses.

There are still unanswered questions. We are currently studying the mechanisms that might lead to productive mathematical sensemaking, intended for a future publication. In addition, the impact on and by the instructors warrants further study. We anticipate that how PITs are implemented is as important as the activities themselves. The effect on student learning associated with instructor fidelity to the PITs' implementation recommendations along with how PITS might be extended to courses beyond the introductory level are areas of future research.

We consider PITs' impact on the instructors' awareness of their own thought processes to be an important outcome of implementing PITs, since instructors are the agents of change who are reframing future discussions and activities to help their students learn. It is not uncommon for instructors who've attended our workshops to create their own PITs once they understand the underpinnings. The main purpose of this paper, and the accompanying web resource, is to make that process more broadly available.

We believe that an intentional instructional focus on student-generated deep mathematical structures, through collaborative quantification and the struggle it entails, has the potential for students to develop mathematical creativity. In this paper, we have presented PITs as one such approach that shows promise in a large enrollment college physics course.

**Acknowledgments**

This work has been supported in part by the National Science Foundation, under DUE #1045250, #1045227, and #1045231. The authors thank Dan Schwartz and Patrick Thompson for advising this project. We thank Eugenia Etkina for valuable feedback on the manuscript and A.J. Richards and Josh Smith for creative contributions to development of the first invention tasks.

## Appendix

### CLASS Scores (2013/2104 combined)

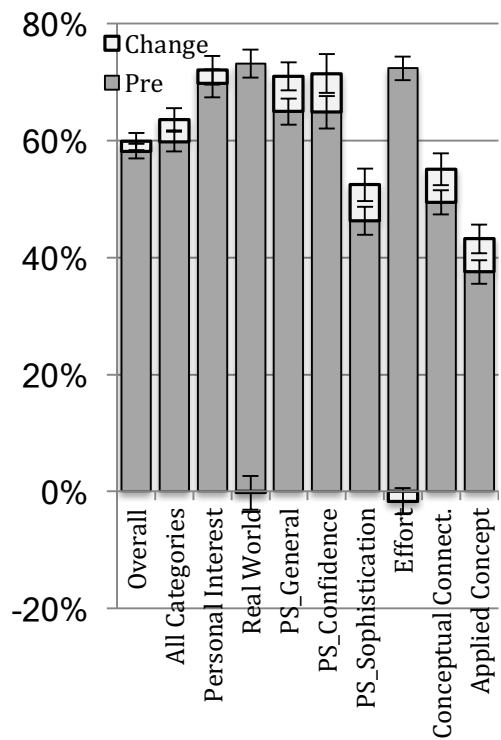

| Category | Pre | Post | Change | Effect size (p-value) |
|---|---|---|---|---|
| Overall | 58.2 ± 1.2 | 59.9 ± 1.5 | +1.7 ± 1.5 | 0.1 n/a |
| All Categories | 59.8 ± 1.7 | 63.6 ± 2.0 | +3.8 ± 1.9 | 0.1 n/a |
| Personal Interest | 69.8 ± 2.4 | 72 ± 2.7 | +2.2 ± 2.5 | 0.1 n/a |
| Real World Connection | 73.1 ± 2.4 | 72.9 ± 2.8 | -0.2 ± 2.9 | 0.0 n/a |
| Problem Solving(PS) - General | 65.0 ± 2.3 | 71.0 ± 2.3 | +6.0 ± 2.4 | 0.2 (.01) |
| PS - Confidence | 64.9 ± 2.8 | 71.5 ± 2.8 | +6.6 ± 3.3 | 0.2 (.05) |
| PS – Sophistication | 46.3 ± 2.4 | 52.5 ± 2.5 | +6.2 ± 2.8 | 0.2 (.03) |
| Sensemak-ing/effort | 72.4 ± 2.0 | 70.7 ± 2.3 | -1.7 ± 2.2 | -0.1 n/a |
| Conceptual Connection | 49.4 ± 2.1 | 55.1 ± 2.5 | +5.7 ± 2.8 | 0.2 (.04) |
| Applied Conceptual Under-standing | 37.5 ± 2.0 | 43.2 ± 2.2 | +5.7 ± 2.4 | 0.2 (.02) |

**Typical quantities invented in the context of college courses:** The mechanics tasks are typically single task activities – with the exception of speed (where the students learn to invent indices) and work (where they first encounter a product quantity). The non-mechanics quantities are more abstract, and tend to involve multi-task sequences.

| Mechanics | | Waves & Thermo | |
|---|---|---|---|
| speed | Ratio | density | Ratio & Product |
| magnitude of acceleration | Ratio | particle flux | Ratio & Product |
| net force (one dimension) | Sum | specific heat | Ratio & Product |
| net force (two dimensions) | Vector sum | latent heat | Ratio |
| spring constant | Ratio | entropy | Ratio |
| coefficient of friction | Ratio | loudness | Logarithm |
| gravitational field | Ratio | **Electricity & Magnetism** | |
| gravitational potential | Ratio | electric potential | Ratio |
| work | Product | electric field | Ratio |
| momentum | Product | electric flux | Ratio |
| impulse | Product | current density | Ratio |
| | | capacitance | Ratio |